\newcommand{\Xcal}{\mathcal{X}}
\newcommand{\xbold}{\mathbf{x}}
\newcommand{\nup}{n_{up}}
\newcommand{\ndown}{n_{down}}
\newcommand{\expectation}{\mathbb{E}}
\newcommand{\fsmooth}{f_{smooth}}

\documentclass[twocolumn]{article}

\usepackage{amsmath, amssymb, amsfonts, amsthm, psfrag, graphicx, subfig}
\usepackage[ruled,vlined]{algorithm2e}

\author{\begin{tabular}[c]{c c c}\normalsize{Firas Hamze} & \normalsize{Neil Dickson} & \normalsize{Kamran
    Karimi}\end{tabular}\\ \normalsize{\emph{D-Wave Systems Inc., 100-4401 Still
      Creek Drive, Burnaby, B.C., V5C 6G9, Canada}}
    \\ \normalsize{\texttt{\{fhamze, ndickson, kkarimi\}@dwavesys.com}} }

\begin{document}

\setlength{\columnsep}{20pt}

\title{{Robust Parameter Selection for Parallel Tempering}}
\date{}

\maketitle
\begin{abstract}
This paper describes an algorithm for selecting parameter values (e.g.
temperature values) at which to measure equilibrium properties with
Parallel Tempering Monte Carlo simulation. Simple approaches to choosing
parameter values can lead to poor equilibration of the simulation,
especially for Ising spin systems that undergo $1^st$-order phase
transitions. However, starting from an initial set of parameter values,
the careful, iterative respacing of these values based on results with
the previous set of values greatly improves equilibration.  Example spin
systems presented here appear in the context of Quantum Monte Carlo.
\end{abstract}

\section{Introduction}
\label{sec:intro}
This paper describes our experience with and modifications to an
algorithm to enhance Parallel Tempering (PT) Monte Carlo
\cite{Hukushima95} known as \emph{Feedback Optimized Parallel
  Tempering} (FOPT) \cite{Katzgraber06}. Though our experimental
analysis will focus on the Quantum Ising Spin Glass in a transverse
field, we shall try to keep the concepts as general as possible as we
believe that workers in as diverse fields as Bayesian Statistics,
Computer Science, and Molecular Simulation may be interested in
optimizing PT on difficult problems. While PT has proved itself to be
an indispensable tool in computer simulation, certain aspects of some
problems can present serious obstacles to its use. The idea of FOPT is
to try to eliminate the barriers in replica-diffusion space that can
hinder the decorrelative effect that PT is supposed to overcome. Our
initial experiments with PT on the Quantum Ising Glass encountered
precisely such difficulties, but we found it necessary to make some
adaptations to FOPT for it to function robustly.

To ease the presentation to those unfamiliar with Quantum Monte Carlo
(QMC) or spin systems but knowledgeable with Markov Chain Monte Carlo
(MCMC, also called \emph{dynamical Monte Carlo}) simulation,
the specialized, physics-oriented concepts are confined to Section
\ref{sec:suzukiTrotter}, which briefly reviews the Quantum Ising Glass
and how it is simulated. Uninterested readers can skip that section and
simply imagine that within is derived a set of intractable
high-dimensional distributions (which in this case are defined over a
binary-valued state space) that we wish to sample from.

Section \ref{sec:partemp} reviews the ideas of PT and FOPT; our
contributions to improving the method are discussed in Section
\ref{sec:FAPT}. Section \ref{sec:suzukiTrotter}, as mentioned above,
discusses quantum spin simulation, and Section \ref{sec:experiments}
presents some numerical data.

\section{Parallel Tempering}
\label{sec:partemp}
\subsection{Basic PT}
\label{sec:partemp:basic}
An excellent general survey of PT can be found in \cite{Iba01}; for
completeness some basic concepts are presented. Suppose we have a
family \( \{ f_{k}( \xbold ) \} \) of \( M \) distributions defined on
a state space \( \Xcal \) each member of which is dependent on some
parameter \( \lambda_{k} \), i.e. \( f_{k}(\xbold) = f( \xbold |
\lambda_{k} ) \). We assume that the terminal parameters \(
\lambda_{1} \) and \( \lambda_{M} \) are fixed and that implementing
naive MCMC at those values, for example local-variable Metropolis or
Gibbs (heat bath) sampling, will result in slowly and quickly
equilibrating (``mixing'') Markov Chains respectively. \( \lambda_{1}
\) and \( \lambda_{M} \) could, for instance, be low and high
temperatures; in the systems we simulate, \( \lambda \) is a parameter
that plays roughly the same role as temperature. The values of the
remaining \( \{ \lambda_{k} \} \) are assumed to be flexible in this
work provided that equilibration gets easier for increasing \( k
\). We refer to a local MCMC simulation at a given parameter value as
a \emph{chain.}

The method of PT supplements the usual within-system Monte Carlo
updates with phases during which swaps of entire states between chains
at different parameter values are attempted. A particular state is
sometimes called a \emph{replica}. In practice, to have a non-negligible
swapping probability, exchanges are only attempted between neighboring
parameters \( \{ \lambda_{k}, \lambda_{k+1} \} \). An exchange is
accepted probabilistically according to the Metropolis ratio
\begin{equation}
\label{eq:alpha}
\alpha = \min \bigg( 1,
\frac{f_{k}(\xbold_{k+1})f_{k+1}(\xbold_{k})}{f_{k}(\xbold_{k})f_{k+1}(\xbold_{k+1})}
\bigg )
\end{equation}
It can be shown that the combination of the local MCMC moves and the
swapping implements a Markov chain on the \emph{joint} state space \(
\Xcal^{M} \) whose invariant distribution is \(
f_{1}(\xbold_{1})f_{2}(\xbold_{2})\ldots f_{M}(\xbold_{M}) \).

The advantage of PT is that in theory, it can enable the system to
escape from local minima it may encounter during the local updates at
large \( \lambda \). It is desirable that a given replica should
drift relatively easily between the terminal \( \lambda \) as this
would suggest that the state at \( \lambda_{M} \) at the end of a
``round trip'' in parameter space has decorrelated from its value at
the start. It may appear at first sight that merely allowing a
reasonable swapping probability would suffice. The latter can be
achieved by spacing the \( \lambda_{k} \) close enough together to
ensure this. One can be more precise (see,
e.g. \cite{Iba01}); it can be shown, by expanding \(
\log f(\xbold|\lambda) \) to second order about \( \lambda_{k} \), that the
expected log acceptance rate between parameters \( k \) and \( k+1 \)
under the equilibrium distributions is
\begin{equation}
\label{eq:logalpha}
\expectation[\log(\alpha)] \sim -I(\lambda_{k})( \lambda_{k+1} - \lambda_{k} )^{2}
\end{equation}
where \( I(\lambda_{k}) = - \sum_{\xbold}
f_{k}(\xbold)\frac{\partial^{2} \log f(\xbold | \lambda )}{\partial
  \lambda^{2}} \big \vert_{\lambda_{k}} \).  We will use this result
in Section \ref{sec:FAPT:initial} to determine the initial spacing for
the feedback procedure and in Section \ref{sec:FAPT:feedback} to
stabilize the algorithm.

\subsection{Feedback-Optimized PT}
\label{sec:partemp:FOPT}

As pointed out by \cite{Katzgraber06}, making PT work properly can be
more subtle than merely assuring sufficient overlap between
chains. Even though the marginal swap rate between all neighboring
parameters may seem reasonable (say \( 25 \% \),) there can exist
higher-order bottlenecks in parameter space that effectively choke the
replica diffusion. A given replica can repeatedly make its way to the
bottleneck but only rarely pass through it. Choosing the \(\{
\lambda_{k} \} \) to remedy this issue is thus not a matter of merely
achieving a given swapping rate between adjacent chains but of
ensuring that a large number of replica round-trips takes place.

We now briefly summarize the insights of \cite{Katzgraber06}. Let \(
\nup(\lambda_{k}) \) and \( \ndown(\lambda_{k}) \) denote, for a given
run of PT, the total number of replica that have visited parameter \(
\lambda_{k} \) that were drifting ``upward'' and ``downward''
respectively. An upward-drifting replica is one that, of the two
terminal parameters, has visited \( \lambda_{1} \) most recently;
a downward-drifting one has last visited \( \lambda_{M} \). Replica that
have not yet visited either endpoint do not contribute to the \( \nup
\) or \( \ndown \) sums. To maximize the rate at which the replica
diffuse from \( \lambda_{1} \) to \( \lambda_{M} \), the flow fraction
\begin{equation}
\label{eq:fraction}
f(\lambda_{k}) = \frac{\nup(\lambda_{k})}{ \nup(\lambda_{k}) +
  \ndown(\lambda_{k}) }
\end{equation}
should decrease linearly from \( \lambda_{1} \) to \( \lambda_{M} \)
\cite{Trebst04}. In \cite{Katzgraber06}, an iterative algorithm is
described, where, after a run of PT with a given set of parameters,
new parameters are generated from the old ones and the measured \( f
\) so as to (hopefully) make \( f \) closer to optimal during the
subsequent run of PT. In other words, the linearly-decreasing \( f \)
is a fixed point of the procedure. The algorithm tends to move the
parameter values away from where \( f \) has a small slope towards
locations where it drops off sharply. For convenience, we restate in
Algorithm \ref{algo:FOPT} a version of it appearing in \cite{Nadler07}
that is somewhat more transparent than the original one defined in
\cite{Katzgraber06}.

\begin{algorithm*}
\DontPrintSemicolon
\SetKwInOut{Input}{Input}
\SetKwInOut{Output}{Output}
\Input{ \( \{\lambda_{k} \} \): Initial parameter set \\
  \( M \): Number of parameters \\
  \( N_{iter} \): Number of feedback iterations \\
  \( N_{sweep} \): Number of sweeps of PT within an iteration
}
\Output{
  \( \{ \lambda_{k} \} \): Optimized parameter set
} 
\Begin{
  \For{\(i=1 \ldots N_{iter}\)} 
  {
    ParallelTempering( \(\lambda_{k} \) ) for \( N_{sweep} \) steps, calculate \( f \)\;
    Define \( g(f) \) such that: 
    \begin{enumerate}
    \item \( g(f(\lambda_{k})) = \lambda_{k} \)
    \item Within \( (\lambda_{k}, \lambda_{k+1}) \), \( g(f) \) is a
      linear interpolation between \( g(f(\lambda_{k})) \) and \( g(f(\lambda_{k+1})) \)
    \end{enumerate}
    \For {\(k=2 \ldots M-1 \)}
    {\( \lambda_{k} \gets g((k-1)/(M-1)) \)}
  }  
}
\caption{Feedback-Optimized PT}
\label{algo:FOPT}
\end{algorithm*}

The procedure can be stopped before the passage of \( N_{iter} \)
iterations if some criterion is specified and met; it may also be
desirable to increase \( N_{sweep} \) from one iteration to the next
\cite{Katzgraber06}.

Impressive results in eliminating a flow bottleneck in the planar
ferromagnetic Ising model are presented in \cite{Katzgraber06}, which
suggested that FOPT was a natural candidate to alleviate a similar
issue we encountered while simulating Quantum Ising glasses with PT.
Unfortunately, in its initial form, the algorithm was unstable on
these problems: one issue was that it was too zealous in its placement
of the parameters around the measured drop-off in the fraction. For a
given number of chains and simulation time, this would cause some
parameter intervals to become too wide, resulting in chains where no
upward \emph{or} downward-moving replica visit and hence breaking the
recursion. We suspect this to be due to the extreme sensitivity of the
systems' behavior around the bottleneck, and the fact that owing to
the large system sizes, it was infeasible to run PT for the times
required to obtain enough statistics to accurately calculate \( f \).
Interestingly, this problem seemed to appear even when the number of
chains was increased dramatically; its occurrence was merely delayed
by a few iterations.

Nonetheless, several factors drove us to make practical modifications
to FOPT, not least of which was the sheer impracticality of manually
choosing good parameter values for a sizable number of very large and
difficult problems. In the next section we detail our implementation
improvements.

\section{Robust Feedback-Optimized PT}
\label{sec:FAPT}

There were several components that we found to be necessary for
feedback-optimization to properly function on our set of problems, but
we emphasize that fundamentally, the goal is the same as that of the
original FOPT, namely achieving a straight-line \( f \).

\subsection{Initial Parameter Spacing}
\label{sec:FAPT:initial}

In some cases, good rules of thumb exist for choosing parameter
spacings in PT; when dealing with temperature, for example, a
geometric sequence has been suggested to be reasonable in certain
circumstances \cite{Predescu04}.  Generally, such guidelines are often
unclear; indeed this is the point of the feedback-optimization idea.
In our practical experience though, the initial parameter spacings and
especially the number of chains can have a substantial influence on
the efficacy of the feedback procedure for a given amount of
computation time. In particular, our experience has shown that
starting with an excessive number of chains can cause poor
performance.

For our initialization strategy, we devised the simple method
AddChains(), described in Algorithm \ref{algo:addchains}. Starting
from a sparse, linearly-spaced initial set of parameters, a short run
of PT takes place within which the exchange rates are
estimated. Parameters are then added in such a way as to achieve some
minimum swap rate.  The number of initial chains can be relatively
small; for example for the size \( N \) spin systems we considered, we
typically used \( \sqrt{N}/4 \) chains (determined heuristically) or
less. When PT is run with such a low number of parameters, the
swapping probability is bound to be very small for quite a few
parameter intervals. In fact if we tried to estimate these swapping
probabilities by histogramming the number of swaps occurring in a
typical-length run, we would probably end up with zeros, especially if
the measurement takes place only after a certain number of
equilibration (sometimes called \emph{burn-in}) steps. However it is
possible, instead of averaging the number of swaps themselves, to
average the log Metropolis ratios between neighboring chains each time
swaps are attempted. Specifically, if during the \( n \)th swap
attempt, we compute the ratio:
\[
\alpha^{(n)}_{k,k+1} = \min\bigg( 1, \frac{f_{k}( \xbold_{k+1} )
  f_{k+1}( \xbold_{k})}{f_{k}( \xbold_{k} )  f_{k+1}( \xbold_{k+1})} \bigg) 
\]
and if \( N_{swap} \) is the number of swap
attempts, we define 
\begin{equation}
\label{eq:ElogAlpha}
\widehat{L_{k,k+1}} = \frac{1}{N_{swap}}\sum_{n=1}^{N_{swap}} \log\big( \alpha^{(n)}_{k,k+1} \big)
\end{equation}

It is not necessary to carry out PT to full equilibrium to usefully
estimate the log swap rates using (\ref{eq:ElogAlpha}); crude
estimates calculated over a relatively short PT run can satisfactorily
inform how many chains need to be added to intervals that are too
wide, even when the estimated swap rates are very low (say \( 10^{-30}
\) or so.)  Determination of the number of needed chains needed in
each interval is then done using the second-order approximation in
(\ref{eq:logalpha}).

A run of the initialization routine would thus proceed as follows:
begin with a minimum swapping threshold \( \alpha_{min} \) (e.g. \( 20
\% \)), and an initial number of chains \( M_{0} \) with \(
\lambda_{M_{0}} = \lambda_{M} \). After measuring the swap rates
resulting from a uniform parameter grid, ``grow'' the number of
chains (to a final value of \( M_{1} \)) attempting to meet the
threshold; practically, the list of new parameters can be generated by
populating it with the initial values, appending the new ones to the
end, and sorting it when finished.

\begin{algorithm}
\DontPrintSemicolon
\SetKwInOut{Input}{Input}
\SetKwInOut{Output}{Output}
\Input{\(M_{0}\): Initial number of parameters\\
\( \lambda_{1}, \lambda_{M} \): Terminal parameters\\
\( \alpha_{min} \): Minimum target swap rate
}
\Output{\( M_{1} \): Number of final parameters\\
\(\{\lambda_{k}\}\): Final set of parameters
} 
\Begin{
  \tcp{Initial parameter list; linear spacing}
  \For{\(k=2 \ldots M_{0}-1 \)}{
    \( \lambda_{k} \) = \( \lambda_{1} +
    k(\lambda_{M}-\lambda_{1})/(M_{0}-1) \)\;
  }
  Run PT with \(\{\lambda_{k}\}\) for \( N_{sweep} \) moves, obtaining \(\{
  \widehat{L_{k,k+1}}\}\) \;
  \tcp{Append parameters if intervals too wide}
  \( M_{1} \gets M_{0} \) \;
  \For{\( k = 1 \ldots M_{0}-1 \)} {
    
    \( R \gets \bigg \lceil \sqrt{
      \frac{\widehat{L_{k,k+1}}}{\log(\alpha_{min})} }
    \bigg \rceil \) \;
    \( \Delta \lambda \gets (\lambda_{k+1} - \lambda_{k})/(R+1) \) \;
    \For{\( j=1 \ldots R \)}{
      \( \lambda_{M_{1} + j} \gets \lambda_{k} + j \Delta \lambda \) \;
    }
    \( M_{1} \gets M_{1} + R \) \;
  }
  
  \( \{ \lambda_{k} \} \gets \mbox{sort} ( \{\lambda_{k}\} ) \) \;
}

\caption{AddChains}
\label{algo:addchains}
\end{algorithm}

The procedure can be repeated with the new parameters,
though we have found this to be unnecessary. Once the initial set has
been determined, the feedback phase, discussed in the next section,
can begin.

\subsection{Feedback Procedure}
\label{sec:FAPT:feedback}
The first problem to tackle in the feedback process is that of
instability due to unreliable \( f \) data or extreme problem
sensitivity. As alluded to in Section \ref{sec:partemp:FOPT}, if \( f
\) possesses a severe drop-off, the original algorithm tends to
over-concentrate the available chains around it, resulting in some
intervals being too wide for any replica to pass through at all in the
next iteration. \( f \) thus becomes mathematically undefined due to
\( \nup=\ndown=0 \), and the algorithm malfunctions. To redress this,
an approach we implemented was to smooth \( f \) in such a way that
the parameters do not move too quickly. We defined \(
\fsmooth(\lambda_{k}) = (1-w)f(\lambda_{k} ) + wL( \lambda_{k}) \),
where \( w \) is a weight between 0 and 1, and \( L( \lambda_{k}) = 1
- (k-1)/(M-1) \), i.e. it decreases linearly from \( 1 \) to \( 0 \)
in \( M \) steps, or alternatively, it is the fixed point of the
optimization algorithm. If \( w=0 \), we recover the original FOPT; at
the other extreme, if \( w=1 \), the feedback does nothing to the
parameters. At intermediate values of \( w \), applying feedback to \(
\fsmooth \) instead of \( f \) has a damping effect on the parameters'
motion. It may be advantageous to reduce \( w \) between subsequent
feedback iterations as the initial estimates of \( f \) become better.
 
In our experiments, the smoothing trick on its own did not always
eliminate the problem of overly-wide intervals, though it certainly
postponed this unfortunate behavior. Consequently, we developed a
method that would override the feedback routine's output if it was
likely that the resultant swap rates were so low that they would cause
the issue discussed above to occur. We called this the
\emph{post-process} step; it is summarized in Algorithm
\ref{algo:postproc}. Using the approximate relation
(\ref{eq:logalpha}) between the log accept rate and the interval
width, this procedure uses the accept rates estimated during the
iteration to predict what the accept rates would be due to the
\emph{new} parameters, i.e. those resulting from the feedback. If, for
a given interval, the predicted rate is lower than a threshold \(
\alpha_{min}\), its width is thresholded. This method may result in
the final interval being too wide, and since the endpoints are assumed
to be static, more parameters are added there to compensate, again
with the objective of attaining at least \( \alpha_{min}\) in the
resultant new intervals. Hence the total number of chains can
continue to grow during the feedback procedure. Of course \(
\alpha_{min} \) need not be the same as the one used in the AddChains
phase; indeed we have found that it should be kept as low as possible
(say around \(5\%\)) in order to minimize intrusiveness on the
parameter search. Its role should be viewed exclusively as one of
``life support.''

\begin{algorithm}
\DontPrintSemicolon
\SetKwInOut{Input}{Input}
\SetKwInOut{Output}{Output}
\Input{\( \{ \lambda_{k} \} \): parameters at start of
  feedback iteration \\ 
  \( \{ \lambda'_{k} \} \): parameters after
  feedback \\
  \( \{\widehat{L_{k,k+1}}\} \): log swap rates measured
  using parameters \( \{ \lambda_{k} \} \) \\
  \( \alpha_{min} \): minimum target swap rate \\
  \(M\): initial number of
\( \{\lambda_{k}\}\)}
\Output{\( \{ \lambda''_{k} \} \): parameters after
  post-process \\
  \( M \): final number of parameters \( \{\lambda''_{k}\}\)}
\Begin { 
\tcp{This stage handles \(\{\lambda_{k}\}\) intervals that are too wide}
\( \lambda''_{1} \gets \lambda_{1} \)\;
\For{\( k = 1 \ldots M-2 \)} {
  \( l \gets \max\{ j | \lambda_{j} \leq \lambda''_{k} \} \)\; \(
  \Delta \lambda \gets \lambda_{l+1} - \lambda_{l} \) \; \( \Delta
  \lambda_{max} \gets \Delta \lambda\sqrt{ \log(\alpha_{min})/
    \widehat{L_{l,l+1}} } \) \; \( \Delta \lambda' \gets
  \lambda'_{k+1} - \lambda_{k} \) \; \If{\( \Delta \lambda' > \Delta
    \lambda_{max} \)} {\( \lambda''_{k+1} \gets \lambda''_{k} + \Delta
    \lambda_{max} \)} \Else {\( \lambda''_{k+1} \gets \lambda'_{k+1}
    \)} } 
\tcp{This stage adds chains to the final interval if needed} 

\( l \gets \max\{ j | \lambda_{j} \leq \lambda''_{M-1} \} \)\;

\( \Delta \lambda \gets \lambda_{l+1} - \lambda_{l} \)\;
 
\( \Delta \lambda_{max} \gets \Delta \lambda\sqrt{
  \log(\alpha_{min})/\widehat{L_{l,l+1}}} \)\;

\( M_{old} \gets M \)\;

\( \lambda \gets \lambda''_{M-1} + \Delta \lambda_{max} \)\;

\While{\( \lambda < \lambda'_{M_{old}} \)} {
  
  \(\lambda''_{M} \gets \lambda \) \;
  
  \( l \gets \max\{ j | \lambda_{j} \leq \lambda \} \)\;
  
  \( \Delta \lambda \gets \lambda_{l+1} - \lambda_{l} \)\;
  
  \( \Delta \lambda_{max} \gets \Delta
  \lambda\sqrt{\log(\alpha_{min})/\widehat{L_{l,l+1}}} \)\;
  
  \( \lambda \gets \lambda + \Delta \lambda_{max} \)\;
  
  \( M \gets M + 1 \)\;

}
\( \lambda''_{M} \gets \lambda_{M} \)\;
} 
\caption{Post-Process Parameters}
\label{algo:postproc}
\end{algorithm}

A somewhat counter-intuitive fact we observed was that in the initial
iterations at least, it can sometimes be favorable to use the
normalized function \( \ndown'(\lambda_{k}) =
1-\ndown(\lambda_{k})/\ndown(\lambda_{M}) \) or its smoothed version
as a surrogate for \( f \). Due to the rapid equilibration at large \(
\lambda_{k}\), this quantity tends to stabilize into its final form
more quickly than \( f \) does; using it instead of \( f \) makes the
implicit assumption that the analogous function \( \nup'(\lambda_{k}) =
\nup(\lambda_{k})/\nup(\lambda_{1}) \) will, when it properly
converges, be symmetric to it, i.e. \( \nup' \approx 1-\ndown'\). This
assumption may end up being incorrect, but it seems in some cases to
be more reliable than using faulty \( f \) data in the beginning. As
with the smoothing of \( f \), as the iterations advance and the
bottlenecks are more accurately located, one can start using the actual \(
f \) in the feedback.

A final implementation aid we noticed to help speed up the
convergence of \( f \) considerably within each feedback iteration was
initializing all the chains with a ground state (global minimum) if it
is known. If the ground state is unique, as it was in the systems we
were considering, this is the correct equilibrium behavior at the
low \( \lambda \) values.

The next section discusses the model we performed our experiments on. It
can be skipped by non-physicists.

\section{Quantum Ising Glass}
\label{sec:suzukiTrotter}

Equilibrium simulation of quantum spin systems can be done by
application of the Suzuki-Trotter (ST) framework \cite{Suzuki76}; the
quantum system's partition function is approximated by a partition
function corresponding to that of a classical Ising model consisting
of multiple ferromagnetically-coupled copies (``Trotter slices'') of
the original system. If one can draw equilibrium samples from this
effective classical system, expectations with respect to the quantum
system can be estimated.  Representations using more slices give a
better approximation but are more computationally demanding to
simulate.
 
Our aim in these simulations was estimation of the minimum energy gap
as a function of the \emph{quantum adiabatic parameter} \( \lambda \in
[0,1] \); the quantum Hamiltonian at a given \( \lambda \) is given by
\( H(\lambda) = A(\lambda)H_{B} + B(\lambda)H_{P} \), where \( H_{B}
\) and \( H_{P} \) are purely quantum and classical Hamiltonian
operators respectively, and the functions \(A(\lambda), B(\lambda) \)
are such that the system is classical and quantum for \( \lambda=0 \)
and \( \lambda=1 \) respectively.
  
Omitting unnecessary details, the Hamiltonians \( H_{P} \) came from
a specific class of non-planar Ising Glasses. Our gap calculation
methodology was the same one appearing in \cite{Young08}, which
crucially depends on obtaining accurate estimates of the spin
correlation function. For large system sizes, na\"ive Monte Carlo
algorithms suffer from equilibration issues at small values of \(
\lambda \), hindering the estimation. It seemed natural that PT would
assist in this problem; the set of target distributions \( f_{k} \)
simply corresponded to the ST representations of the quantum systems
at the \( \{\lambda_{k}\} \). It turned out that for many such
problems, PT can have serious problems of the sort discussed in this
paper; their presence corresponded to the existence of first-order
quantum phase transitions. Tuning the parameters by hand is very
difficult for such problems; small changes to the parameters had
dramatic effects on \( f \). A PT optimization method was thus
essential for simulations we were interested in. As an
added benefit, the method we implemented tends to concentrate the
parameters precisely around the point we were most interested in,
namely where the spectral gap is minimum.

\section{Experiments}
\label{sec:experiments}

The PT optimization algorithm discussed in this paper was used to
determine parameter spacings for hundreds of quantum spin systems,
which, in their approximate representation, correspond to complex,
binary-valued systems of up to \( 32768 \) variables (these resulted
from representing a 128 quantum spin system with 256 Trotter slices.)
The parameters were determined individually for each instance as it
was observed that a set optimized for one instance was likely to
perform poorly on other instances in the same class of problems.

For stepwise illustration, let us consider a particular 16-spin
Quantum Ising Glass represented with 128 Trotter slices to yield a
system size of \( N = 2048 \). This specific instance was extremely
difficult to simulate with PT owing to a sharp first-order quantum
phase transition. If the parameters were spaced so as to achieve a
uniform swap rate, a dramatic flow bottleneck manifested itself. It
was also impervious to the initial FOPT algorithm, as well as numerous
other approaches which seemed promising on other problems.

In order to allow a fair comparison, in both the original FOPT and in
the version with our improvements, we applied the AddChains
initialization routine to determine the starting parameters, and the
states were all seeded with the ground state.

\begin{figure*}[hbpt!]
\centering
\includegraphics[width=0.5\textwidth]{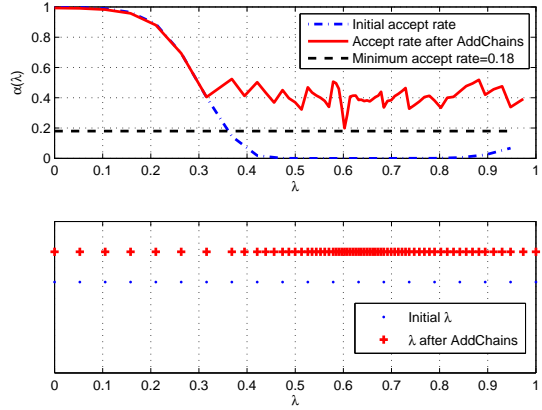}
\caption{(Color online) Illustration of the AddChains algorithm on an
  \( N=2048 \) Ising spin system. In the bottom graph, we show the
  grid of initial linearly-spaced parameters and the resultant \(
  \lambda \) after AddChains. The top graph plots the initial and
  final estimated swap rates after \( 70000 \) sweeps of PT (in the
  left side of the plot, the blue and red lines are superimposed.)
  Around the region with \( \lambda \in [0.5,0.8] \) the initial
  estimated swap rate plunged as low as \(\approx e^{-79}\); after the
  parameter addition, all intervals had swaps taking place over the
  criterion rate, shown by the black dotted line. The initial grid
  consisted of \( 20 \) parameters, to which 46 were added by the
  algorithm.}
\label{fig:addchains}
\end{figure*}

First we demonstrate how Algorithm \ref{algo:addchains}
performed. AddChains was called on this system using a target swap
threshold of \( 18 \% \), \( 70 000 \) sweeps of PT, and \( 20 \)
chains to start. Figure \ref{fig:addchains} shows both how the
acceptance rates responded to the interspersion and the particular
locations in \( \lambda \) space that parameters were added. The
estimator of the swap rates at the initial parameter
spacing produced \( \widehat{L} \) as low as \(-130\); thus in
the \(70000\) sweeps we performed, actually observing any swaps in
such an interval is virtually impossible. The lower plot in Figure
\ref{fig:addchains} shows where parameters were added to achieve the
target; note the concentration around \( \lambda \approx 0.6 \). Our
rough algorithm succeeded in achieving the target rate in all
intervals.

\begin{figure*}[htbp!]
\centering
\subfloat[]{\includegraphics[width=1.2\columnwidth]{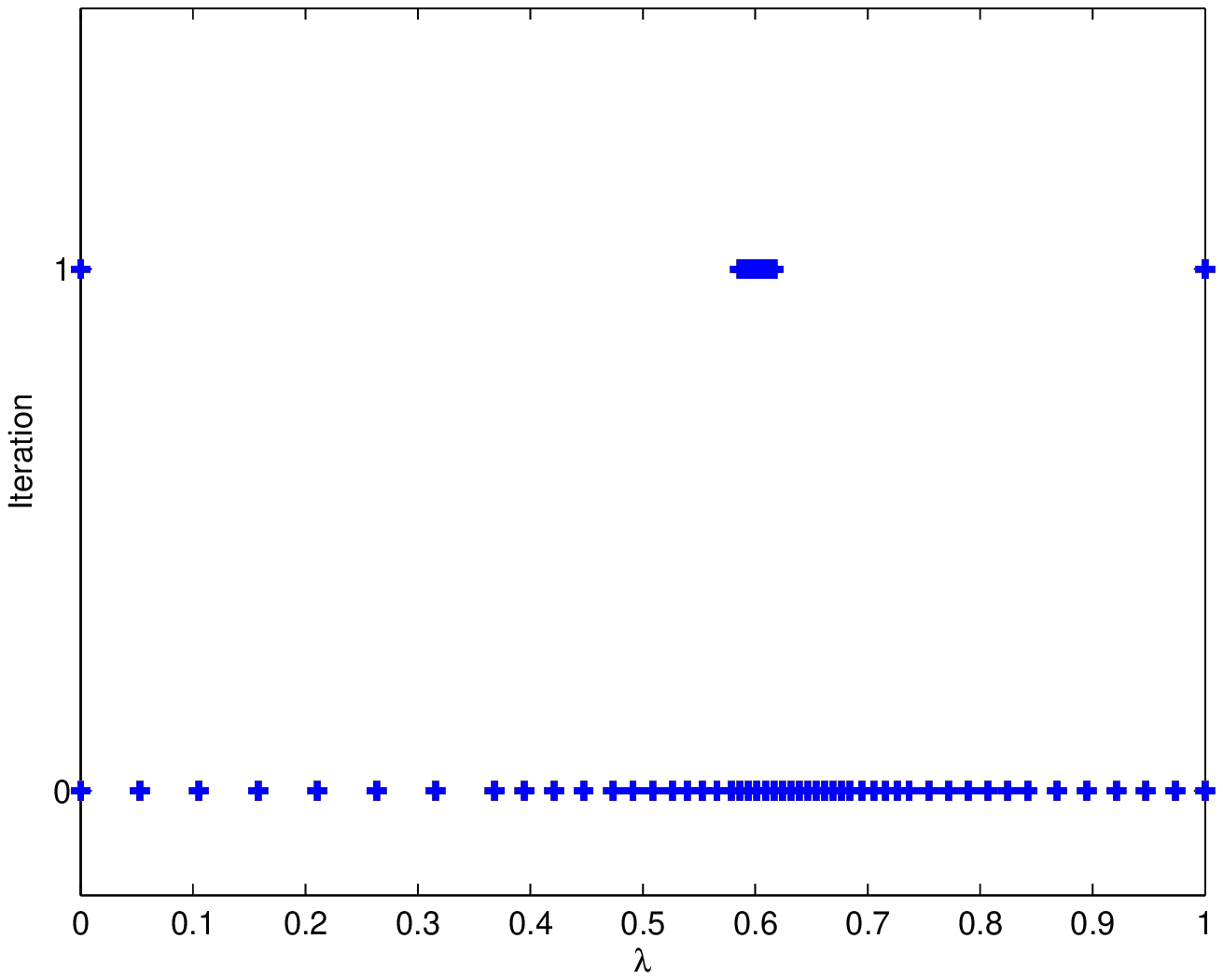}}\\
\subfloat[]{\includegraphics[width=1.2\columnwidth]{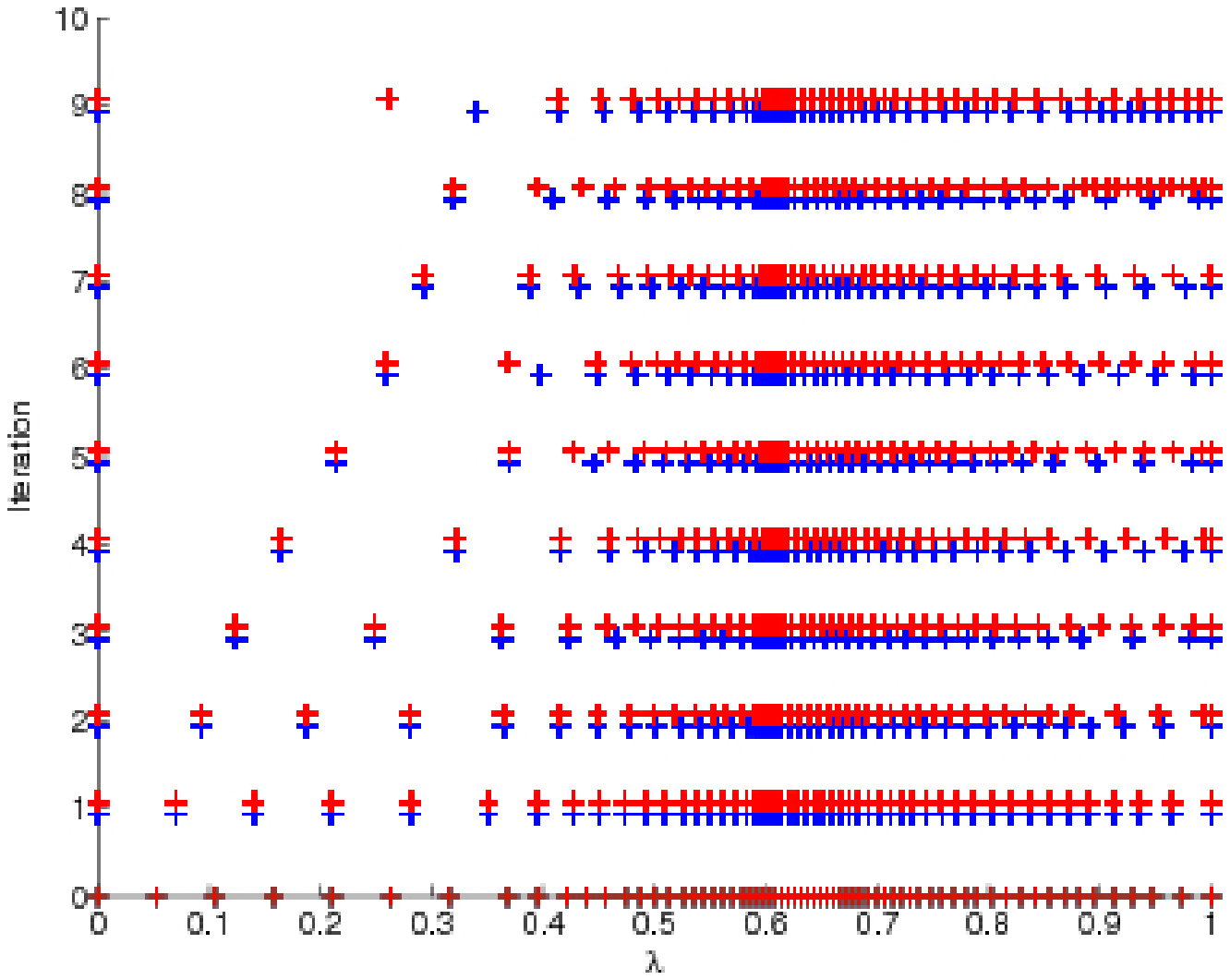}}
\caption{(Color online) Illustration of the progression of the parameters using the
  original FOPT algorithm and the one with our modifications on the
  \(N = 2048 \) Ising spin system discussed in the text. Both
  algorithms began with the \( \lambda \) spacings generated by
  AddChains; all replica were initialized with the ground state. At
  left, we see that in a single iteration, FOPT spaces all the
  parameters around \( \lambda \approx 0.6 \), resulting in very wide
  intervals between the endpoints. The algorithm could thus not
  proceed past this point because the replica in the middle could no
  longer visit either terminus; consequently \( f \) was
  mathematically undefined. Our algorithm's output appears at
  right. At each iteration step the resultant \( \lambda \) using a
  smoothed version of \( f \) in FOPT is shown in blue; the \( \lambda
  \) returned when those parameters are passed to the PostProcess
  algorithm appear in red. Note that to keep the swap rate over the
  threshold of \( 3 \% \), PostProcess often ``pulled'' parameters back
  if it predicted that the interval was too wide.}
\label{fig:foptfaptrun}
\end{figure*}

\begin{figure}
\centering
\includegraphics[width=\columnwidth]{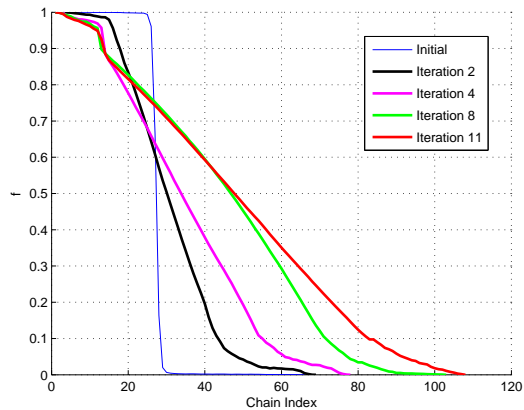}
\caption{(Color online) Progress of the upward-moving fraction \( f \)
  when our algorithm is applied to the \( N=2048 \) Ising spin system
  discussed in the text. Note the abrupt decline when PT is run with
  the results of AddChains. Subsequent iterations progressively make
  \( f \) closer to the desired form, namely a straight line. \( 200
  000 \) PT sweeps were used to gather the \( f \) statistics at each
  iteration.}
\label{fig:FAPTfProgress}
\end{figure}

In Figure \ref{fig:foptfaptrun} a run of the initial FOPT algorithm
and our stabilized version are shown. For our algorithm, we used the
fairly strong smoothing value of \( w = 0.75 \); the number of PT
sweeps was constant at \( 200 000 \) in each iteration. For
PostProcess, the minimum interval swap rate was set to \( 3 \% \). The
details can be read from the caption in the Figure, but the essential
point is that the original algorithm aggressively spaces the
parameters in such a way that the recursion cannot proceed past a
single iteration. For all but the endpoint parameters, a PT run of
realistic length using such \( \{\lambda\} \) would give \( \nup =
\ndown = 0 \). For illustration, in Figure \ref{fig:FAPTfProgress} we
show the progression of \( f \) (not \( \fsmooth \) ) at different
iterations of our method. Even on this highly troublesome instance,
the parameters were finally spaced in such a way that enough
round-trips took place to allow accurate estimation of the minimum
excitation gap.

\begin{figure}
\centering
\includegraphics[width=\columnwidth]{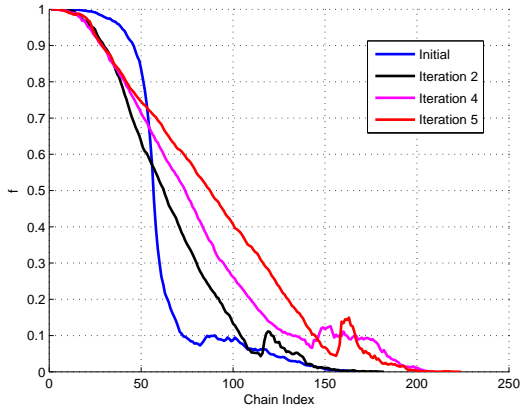}
\caption{(Color online) Progress of the upward-moving fraction \( f \)
  when the algorithm is applied to the \(N=24576\) Ising spin system
  discussed in the text.}
\label{fig:FAPTfBigSystem}
\end{figure}

To demonstrate efficacy on the typical system sizes we were ultimately
interested in simulating, in Figure \ref{fig:FAPTfBigSystem} we
present results of the evolution of \( f \) on an \( N=24576 \) spin
model resulting from representing a \( 96 \) spin quantum system with
256 Trotter slices. The results were obtained using the same
simulation parameter settings as those used for the
previously-discussed 16-spin system. The resultant parameter spacing
again allowed sufficiently good PT performance for us to obtain
high-quality Monte Carlo data.

\section{Conclusion}
\label{sec:conclusion}
We have presented improvements to the problem of iterative parameter
selection for PT. Our contribution has consisted of three parts:
first, an initialization strategy to place parameters in a reasonable
manner when no \emph{a priori} information is known about the problem
domain; second, the introduction of damping mechanisms to the FOPT
algorithm of \cite{Katzgraber06} that assist in tackling the
problem of instability; third a post-processing procedure that
prevents the algorithm from malfunctioning. We demonstrated the
effectiveness of the method by showing experiments on a particularly
difficult quantum spin system represented as a classical Ising model
with \( 2048 \) spins. We have tested our algorithm on much bigger
systems with satisfactory results.

\subsection*{Acknowledgements}
We would like to thank Helmut Katzgraber, Peter Young, Mohammad Amin,
and Geordie Rose for valuable discussions and insights.

\bibliographystyle{plain}
\bibliography{FirasBibliography}

\end{document}